\documentclass[aip,reprint]{revtex4-1}
\usepackage[draft]{hyperref} 
\usepackage{amsmath}
\usepackage{amsfonts}
\usepackage{graphicx}
\usepackage{subfigure}
\usepackage{float}

\begin{document}

\title{An effective cavity resonance model for enhanced optical transmission through arrays of subwavelength apertures in metal films}

\author{Eli Lansey*}
\affiliation{Department of Physics, Graduate Center and City College of the City University of New York}
\email{elansey@gc.cuny.edu}
\author{Isroel M. Mandel}
\affiliation{Department of Physics, Graduate Center and City College of the City University of New York}
\author{Jonah N. Gollub}
\affiliation{Phoebus Optoelectronics, LLC, New York, NY}
\author{David T. Crouse}
\affiliation{Department of Electrical Engineering, City College of the City University of New York}

\begin{abstract}
We present a novel theoretical approach for modeling the resonant properties of transmission through subwavelength apertures penetrating metal films. We show that cavity mode theory applies to an effective resonant cavity whose dimensions are determined by the aperture's geometry and the evanescent decay lengths of the associated diffracted waves. This method suggests a concrete physical mechanism for the enhanced transmission observed in periodic aperture arrays, namely it is the evanescently scattered light, localized in the near field of metal surface, which couples into the apertures. Furthermore, it analytically predicts the frequencies of peaks in enhanced transmission, the quality factor of the peaks, and explains their dependence on variation in the hole radius, periodicity, and the film thickness over a wide range of geometries. This model demonstrates strong correlation to simulation and existing results with a high degree of accuracy.
\end{abstract}


\maketitle

\section{Introduction}
Enhanced optical transmission (EOT) through a periodic array of subwavelength apertures in a metal film was first reported by Ebbesen in 1998\cite{EbbesenNature1998}. Since then, many theories have been proposed for the underlying physical mechanism for this effect. Some have tried to generalize Bethe's theory for a single, extremely small hole through an infinitely thin perfect electric conducting (PEC) film,\cite{BethePR1944} to account for periodic arrays,\cite{GordonPRA2007} and finite thicknesses.\cite{Garc'iadeAbajoPRL2005} However, these approaches are limited to extremely small holes and perfect conducting metals, two situations which are not typically realized.

To avoid the geometric and material limitations of approaches based on Bethe's theory, one common argument supposes that it is the excitation of surface plasmons (SP) on the periodic surface, both for one-dimensional gratings or two-dimensional aperture arrays, which allows coupling of light from an incident wave through the holes.\cite{GhaemiPRB1998,KrishnanOC2001,PendryS2004,CrouseOE2005a,Garcia-Vidal2010} This approach explains the role that structure periodicity plays in EOT. However, it does not account for the variation in EOT due to aperture shape.\cite{KoerkampPRL2004,MolenPRBCmamp2005} Furthermore, EOT has been demonstrated with PEC structures, as well as non-metallic materials where SP contributions are nonexistent.\cite{ThioJOSAB1999,SarrazinPRE2003,LezecOE2004}

Another approach describes the contribution to EOT of propagating waveguide modes along the aperture.\cite{GordonOE2005,ShinPRB2005,RuanPRL2006,CollinOE2007,LanseyJOSAB2011} These arguments suppose that incident light can only propagate through a film and contribute to EOT in a manner which satisfies a waveguide condition along the length of the aperture. These approaches capture some limitations that the individual cavity structures place on allowed frequencies which demonstrate EOT, but do not directly explain the effect of periodicity on EOT or predict the locations of specific peaks in EOT and their dependence on cavity shapes.

This effect has also been studied extensively through use of a semi-analytical coupled wave analysis,\cite{MorenoOE2004,SturmanPRB2008,Garcia-Vidal2010,DelgadoOE2010,GorkunovPRB2011} as well as by many different finite element or finite difference numerical simulation approaches,\cite{PapadopoulosAO2009,BendoymPaSoODX2011} and experimentally.\cite{GhaemiPRB1998,KrishnanOC2001,KoerkampPRL2004,PendryS2004,CrouseOE2005a,Garcia-Vidal2010} These methods all empirically shed light on the dependence of EOT on structure periodicity and cavity shapes, but do not provide an intuitive, or fundamentally clarifying approach toward the mechanisms of EOT. Furthermore, these approaches are often computationally and experimentally expensive to carry out.

Thus, a complete first-principles approach to explain the effect of EOT through two-dimensional arrays of subwavelength holes in metal films is desirable. In this paper we develop an approach which analytically and intuitively explains the physical mechanism of EOT, and completely explains the aforementioned dependence on structure periodicity and cavity shape. The theory is accurate over an extremely broad range of geometrical configurations. In this approach, we extend the idea of waveguide dispersion analysis to account for finite film thicknesses.

For a finite film, an impedance mismatch between the superstrate, i.e.\ the material above the film, and the metal at the top and bottom of the cavities introduces a restriction on the wavelengths that exhibit resonant behavior along the $z$-direction. It is at these resonances, where light is strongly coupled into and through the apertures, where peaks in EOT are manifest. There have been some successful studies of cavity-type resonances for one-dimensional gratings, under limited geometrical conditions;\cite{TakakuraPRL2001,Garcia-VidalPRB2002,GorkunovPRB2011,Guillaum'eeOE2011} we extend this approach to two-dimensional arrays and a larger range of geometries.

Our goal is to describe an effective resonant cavity which has resonant properties that match that of the actual aperture array. It should be emphasized that this is not an \emph{actual} cavity resonance, i.e.\ the fields do not demonstrate standing-wave behavior and there is a flow of energy along the aperture, but an effective cavity resonance (ECR) where the physical extents of the equivalent cavity are determined by the material's structural and material properties.

\section{Effective cavity resonance solutions}
\begin{figure}[h]
    \centering
\subfigure[ A top-down view of the structure under consideration.]{\includegraphics[width=.45\linewidth]{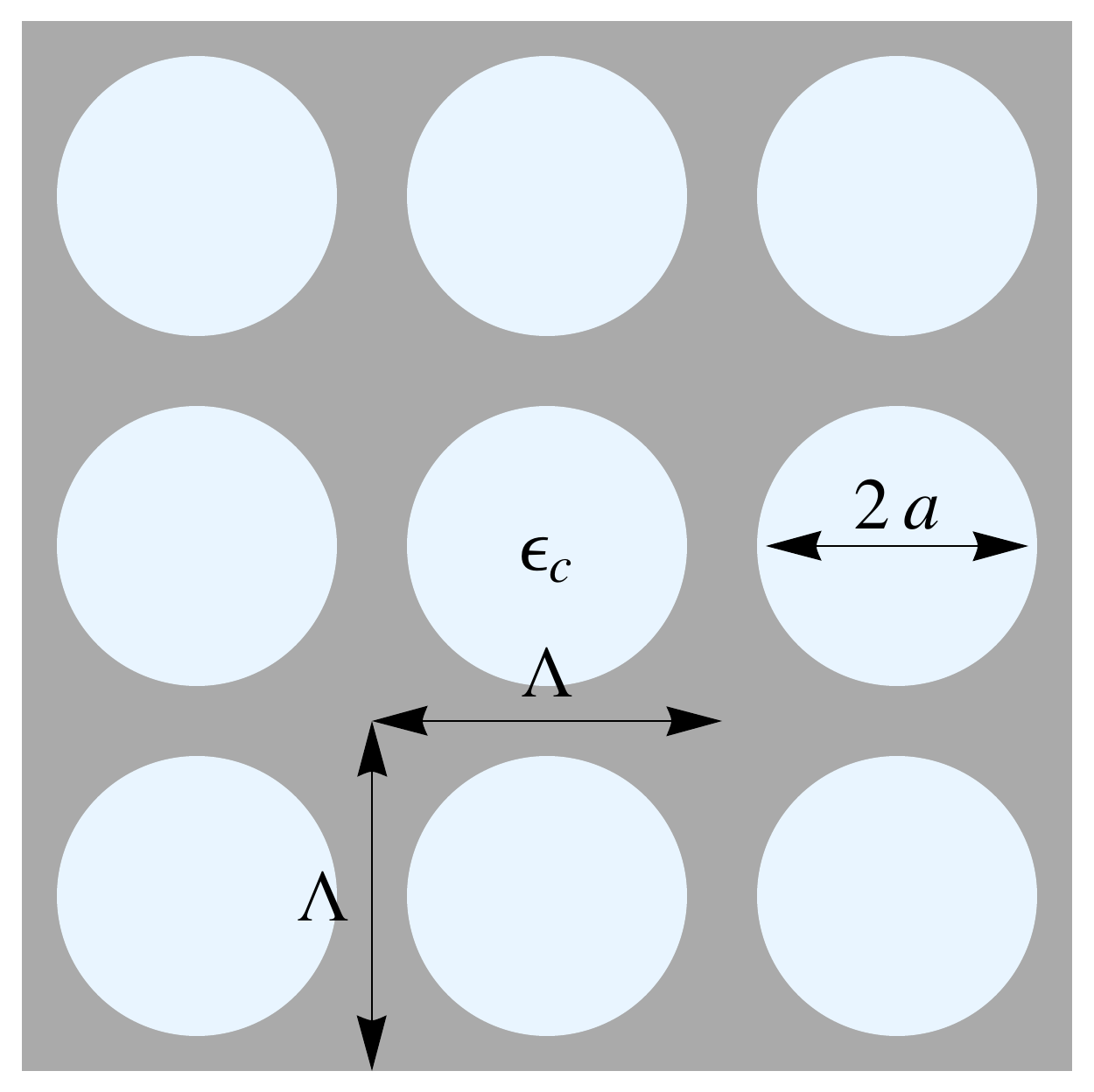}
    \label{fig:topview}
    }\hfill
    \subfigure[A cross section view of the structure under consideration.]{\includegraphics[width=.45\linewidth]{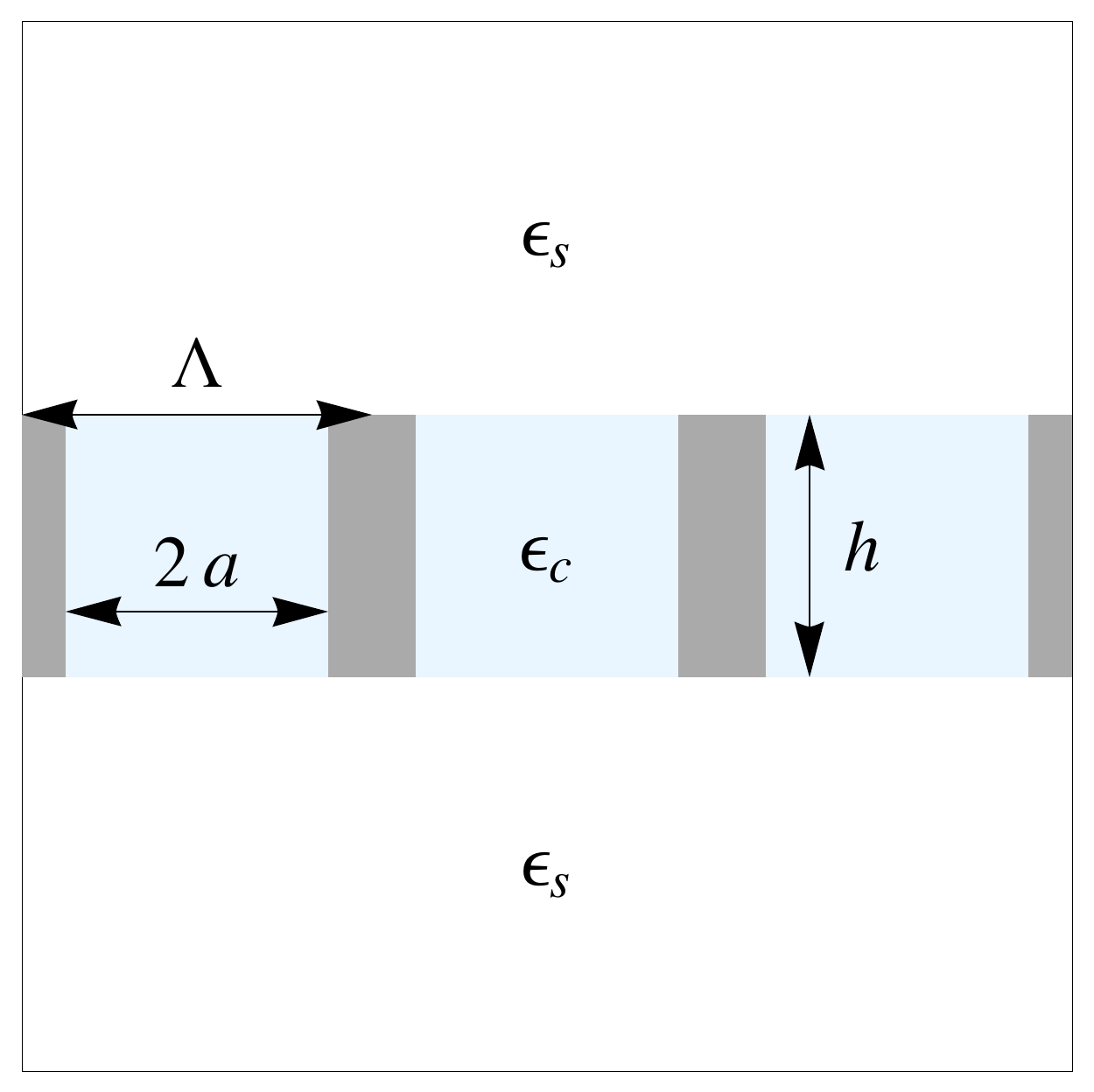}
    \label{fig:sideview}
    }
    \caption{A schematic of periodic cylindrical channels in a thin film is shown from top down \subref{fig:topview} and in cross section \subref{fig:sideview}. The gray region represents the metal, the light blue regions are the dielectric-filled apertures, and the white is the superstrate and substrate.}
    \label{fig:cylinder}
\end{figure}

Here we discuss cylindrical apertures of radius $a$, filled with dielectric $\epsilon_c$ embedded in a metal film of thickness $h$, arranged in an infinite square periodic lattice of period $\Lambda$, with a dielectric, $\epsilon_s$, above and below the film, see Fig.~\ref{fig:cylinder}. The approximation of an infinite lattice is valid in practice as long as the size of an complete hole array is significantly larger than the wavelength of incident light, where we can neglect edge effects.\cite{Ashcroft1976} We additionally neglect any magnetic effects, taking $\mu=1$ for all materials, and assume an implicit $\exp[-i\omega t]$ harmonic time dependence.

%
%

Considering an infinitely-long aperture, the dispersion relation of light within these structures is
\begin{equation}
\label{eq:disp}
\epsilon_c\frac{\omega^2}{c^2}=k_z^2+\beta^2,
\end{equation}
where $\omega$ is the angular frequency, $c$ is the speed of light, and $k_z$ is the propagation wavevector. The transverse wavevector $\beta$ is found by evaluating boundary conditions at the cavity's metal walls. For apertures embedded in a PEC film, its values are
\begin{subequations}
\label{eq:PECbeta}
\begin{align}
\beta_0^{TM} &\equiv \chi_{mn}/{a},\\
\beta_0^{TE} &\equiv \chi\,'_{mn}/{a},
\end{align}
\end{subequations}
where $\chi_{mn}$ is the $n$th root of the $m$th Bessel function of the first kind, and $\chi\,'_{mn}$ is the $n$th root of the derivative of the $m$th Bessel function of the first kind. Using the skin depth boundary condition (SDBC) for realistic metals, we have
\begin{equation}
\label{eq:SDBCbeta}
\beta = \left(\frac{1}{1+\xi}\right)\beta_0,
\end{equation}
where $\xi\equiv\delta_m/a$, and $\delta_m$ is the skin depth in the metal. This approach is discussed in detail by Lansey et al.\cite{LanseyJOSAB2011} Note that Eq.~\eqref{eq:disp} captures the dependence of the cavity shape and dielectric, as well as metal properties, but does not yet account for any properties of the superstrate or periodicity.

The remainder of this work involves determining an appropriate restriction on $k_z$ due to the finite film thickness and periodicity of the structure. If the restriction forces $k_z$ to take discrete values, it changes the allowed $\omega$ in Eq.~\eqref{eq:disp} from a smoothly varying range of values to distinct resonance frequencies. Again, we note that this ECR is not an actual cavity resonance, but there are still spatial restrictions which introduce a buildup in field strength within the apertures which can be modeled as an effective resonant cavity.

\subsection{The Fabry-Perot model}
\label{sec:illus}
We first investigate a simple, Fabry-P\'erot (FP) model for a restriction on $k_z$ which sets up a resonance condition, which we will study in greater detail in Section.~\ref{sec:ecr}, and serves to illustrate our general approach. It is worthwhile noting that, due to its simplicity of form, this first approximation is regularly cited when discussing theory and design of aperture array materials.\cite{HibbinsPRL2004} Here the waveguide has a finite height, $h$, and we assume that this distance sets the resonance condition, whence,
\begin{equation}
\label{eq:kzNoSD}
k_z=p\,\pi/h,
\end{equation}
where $p$ is an integer.

\begin{figure}
    \centering
        \includegraphics[width=1.0\linewidth]{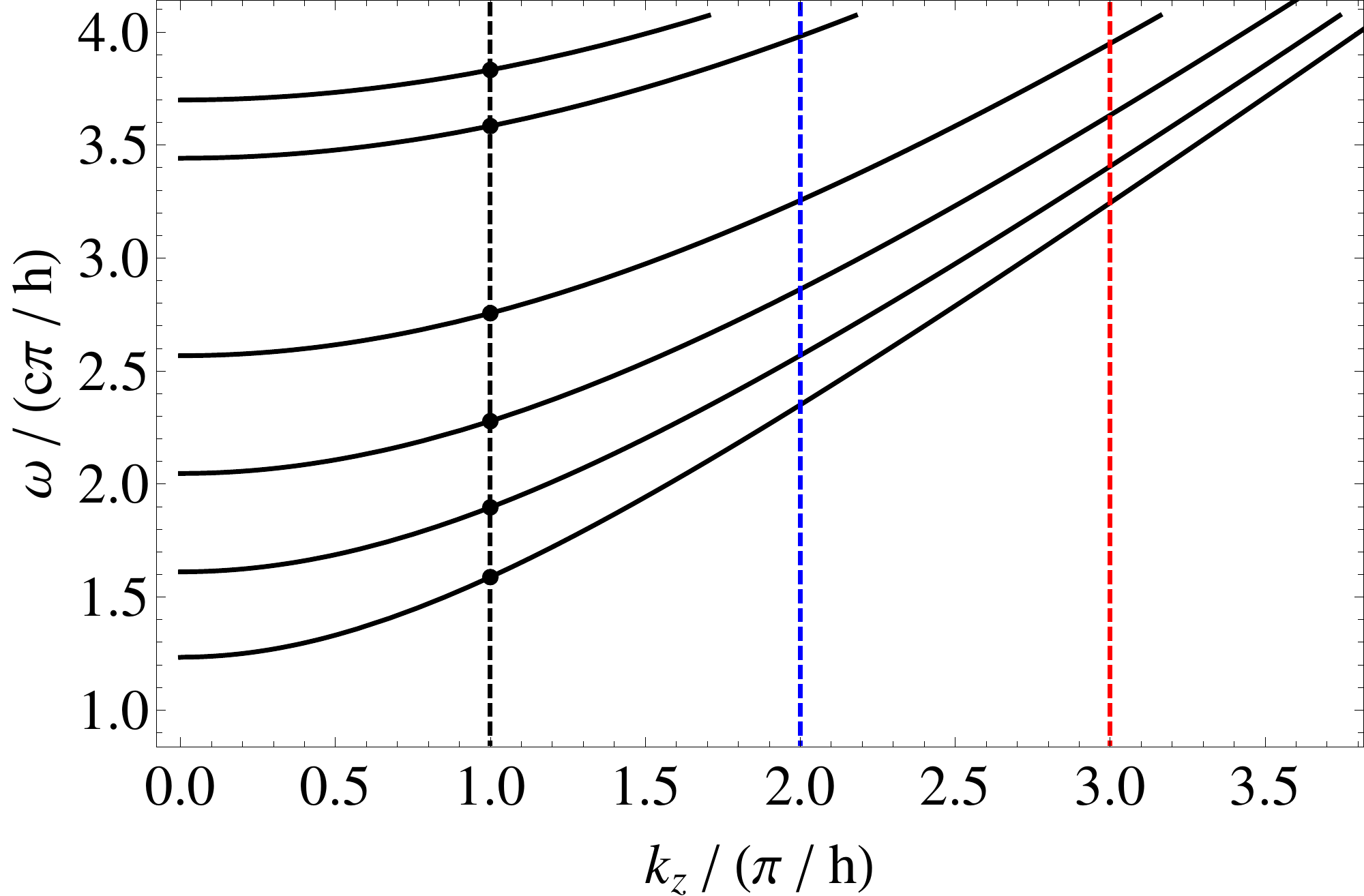}
    \caption{A graphical interpretation of the resonance condition described by Eq.~\eqref{eq:resonanceNoSD}. Vertical lines correspond to the $k_z$ restricted by film thickness (Eq.~\eqref{eq:kzNoSD}), solid curves are the modal dispersion curves of the cylindrical cavities (Eq.~\eqref{eq:disp}). The intersections of the two curves (points for $p=1$ line) correspond to resonance conditions. Here $h/a=2.1$.}
    \label{fig:kzSolveNoSD}
\end{figure}

Substituting this constrained value for the propagation constant into Eq.~\eqref{eq:disp} gives a set of discrete resonance frequencies,
\begin{equation}
\label{eq:resonanceNoSD}
\omega_{mnp}=\frac{c}{\sqrt\epsilon_c}\left[\left(\frac{p\,\pi}{h}\right)^2+\beta_{mn}^2\right]^{1/2}.
\end{equation}
Fig.~\ref{fig:kzSolveNoSD} shows a graphical interpretation of this method. The dispersion curves are plotted, along with the restricted values for $k_z$ from Eq.~\eqref{eq:kzNoSD} which are vertical lines. The intersections between these curves correspond to resonance conditions for the effective cavity. 
%

\begin{figure}[h]
    \centering
        \includegraphics[width=.45\linewidth]{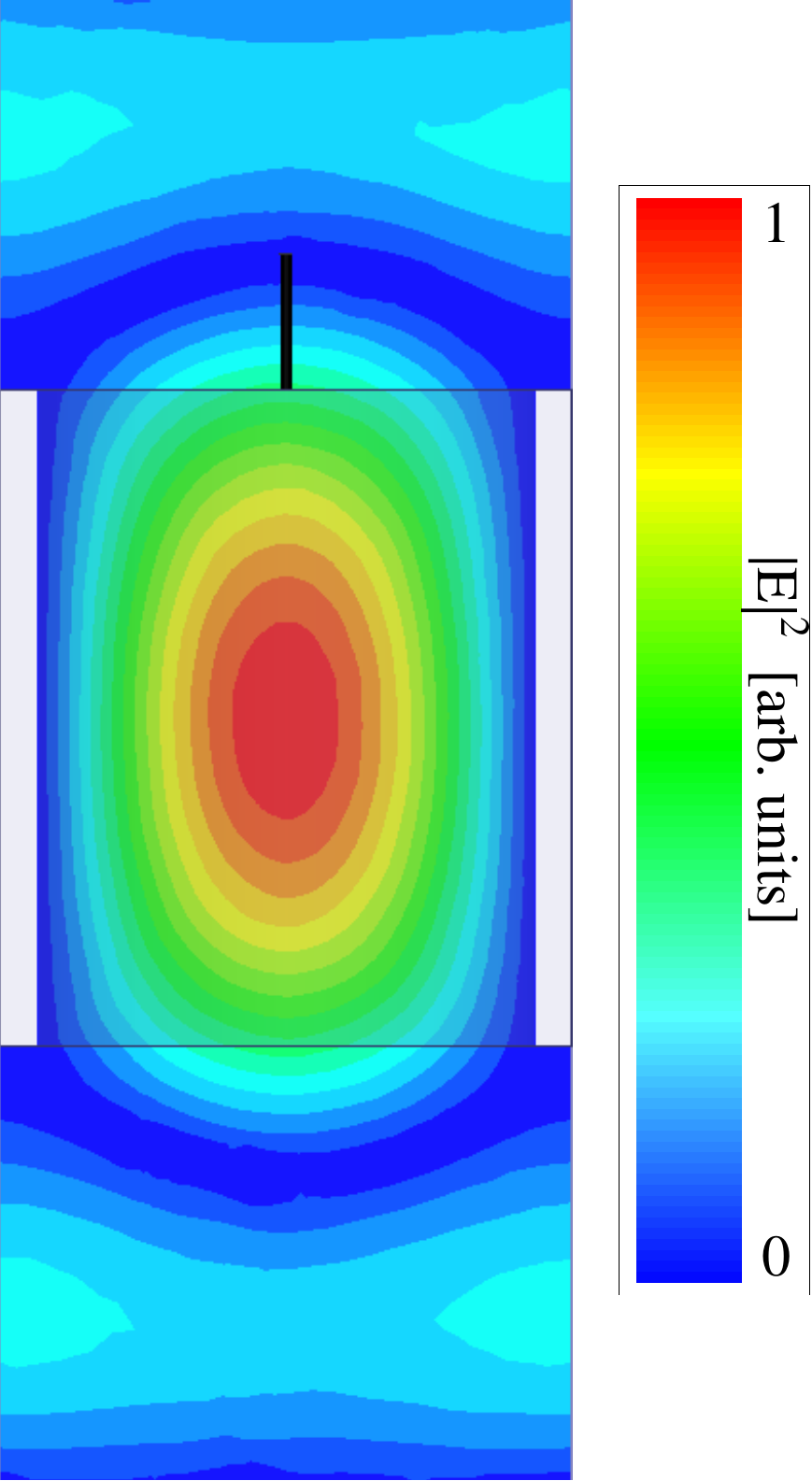}
    \caption{A cross-sectional slice of unit cell of a cylindrical cavity embedded in a PEC film is shown, with the magnitude of the electric field (in arbitrary units) plotted. Here $2a/\Lambda=0.88$, $h/a=2.63$ and the fields are evaluated at $\omega/(c\pi/h)=1.72$. The solid black line, oriented along the $z$-direction above the film surface, is the predicted cavity field leakage depth.}
    \label{fig:field}
\end{figure}

This FP model is insufficient for describing many of the effects of EOT.\cite{TakakuraPRL2001,CaoPRL2002,Garcia-VidalPRB2002,SturmanPRB2008} This solution does not depend on periodicity or the superstrate dielectric and ignores the contribution of the incident fields. Additionally, full field simulations (Fig.~\ref{fig:field}) of aperture arrays show that the fields at a transmission peak reach beyond the surface of the film, above and below, further challenging the notion of using $h$ as the FP cavity height.

\subsection{A complete solution}
\label{sec:ecr}
We are still able to use the resonance condition \eqref{eq:resonanceNoSD} by simply extending the effective height of the cavity. That is, if the fields extend a distance $\delta_e$ above and below the metal surface, the aperture has an effective height
\begin{equation}
\label{eq:effHapt}
h_{\textrm{eff}}=h+2\delta_e,
\end{equation}
which is the actual aperture height plus the total penetration depth into the superstrate and substrate, beyond which the cavity fields decay to zero. Note, that the approach of finding effective heights of cavities has been demonstrated in a circuit model for EOT in some grating structures.\cite{MedinaMTaTITo2010} Then, we substitute this effective cavity height in Eq.~\eqref{eq:kzNoSD}, giving the new restriction
\begin{equation}
\label{eq:kzYesSD}
k_z=p\,\pi/h_{\textrm{eff}},
\end{equation}
where $p$ is an integer. The distance the fields leak out of the cavity is determined by restrictions on the fields above and below the film, which depend explicitly on the periodicity of the apertures.

To find this distance, we must find the maximal spatial extent, in the $z$-direction, of localized fields in the superstrate. Here, we develop an approach which describes the role of evanescent, scattered diffracted fields in the superstrate in setting the spatial extent of the cavity. These localized fields, unlike incident plane waves, are able to couple to waveguide modes in the cavity.

We note that, due to continuity boundary conditions between an aperture and the superstrate, single-walled aperture structures are unable to support the normally-incident TEM waves which strike them. Specifically, TEM waves have no $z$-component to their fields, while the fields in the apertures require a non-zero $E_z$ (TM modes) or $H_z$ (TE mode).\cite{Jackson2} Likewise, light exiting an aperture can not directly excite a zero-order transmission plane wave. Nevertheless, normally-incident light scatters from periodic arrays of subwavelength holes in a manner which satisfies Bloch's theorem, with a spatial dependence of
\begin{equation}
\exp\left[i\left(k_{xm}x+k_{yn}y + k^f_{zmn}z\right)\right],
\end{equation}
where
\begin{subequations}
\label{eq:ksup}
\begin{align}
k_{xm} &=m K\\
k_{yn} &=n K\\
\label{eq:kzFq}
k^f_{zmn} &=\sqrt{\epsilon_s\frac{\omega^2}{c^2}-(m^2+n^2)K^2}
\end{align}
\end{subequations}
where $m$ and $n$ are integers, and $K=2\pi/\Lambda$ is the reciprocal lattice vector.\cite{Ashcroft1976} Note that this $k^f_{zmn}$ is not, in general, the propagating cavity $k_z$ used earlier.

These evanescent diffracted modes, where $(m^2+n^2)K^2>\epsilon_s\omega^2/c^2$, introduce strong localized fields above and below the metal film. These fields decay exponentially away from the surface with a decay length
\begin{equation}
\delta_{mn}\left(\omega,\Lambda\right)=1/\mathrm{Im}[k^f_{zmn}(\omega,\Lambda)].
\end{equation}
In general, these scattered fields have non-zero $z$-components, which then interact with the fields inside the apertures. Thus, the strength, and scattered, although localized, nature of the evanescent modes is what drives the transmission through the film.

The maximal spatial extent of these localized fields then sets the penetration depth of the cavity fields into the superstrate and substrate. That is,
\begin{equation}
\delta_e(\omega,\Lambda)=\max\left[\delta_{mn}\left(\omega,\Lambda\right)\right].
\end{equation}
The inclusion of the $\max$ function picks the mode with the longest decay length which is, in general, the lowest-order non-propagating mode. Although the detailed behavior of the electromagnetic fields above and below the metal is due to a superposition of all evanescent and propagating diffracted modes, the \emph{extent} of the localized fields is still limited by the lowest-order evanescent diffracted mode in the superstrate or substrate. This sets the maximal distance over which any evanescent diffracted modes can extend, and thereby the effective depth of the fields above and below the film. This predicted length is shown in Fig.~\ref{fig:field} as the solid black line, oriented along the vertical direction above the film surface, and is seen to match the field simulation. Additionally, note the field strength at that point matches the field strength at cavity walls, and thus describes the edge of the effective cavity.

This length depends explicitly on the periodicity, which directly explains both the enhancement seen for periodic structures, and the variation in EOT with changes of periodicity. Furthermore, the dependence on superstrate dielectric is also explained in this approach, as it affects the diffracted modes. These effects are captured by Eq.~\eqref{eq:kzFq}. Thus, we have a resonance condition which depends on all relevant structural properties. We study this further in Section~\ref{sec:geom}.

\begin{figure}[h]
    \centering
        \includegraphics[width=1.0\linewidth]{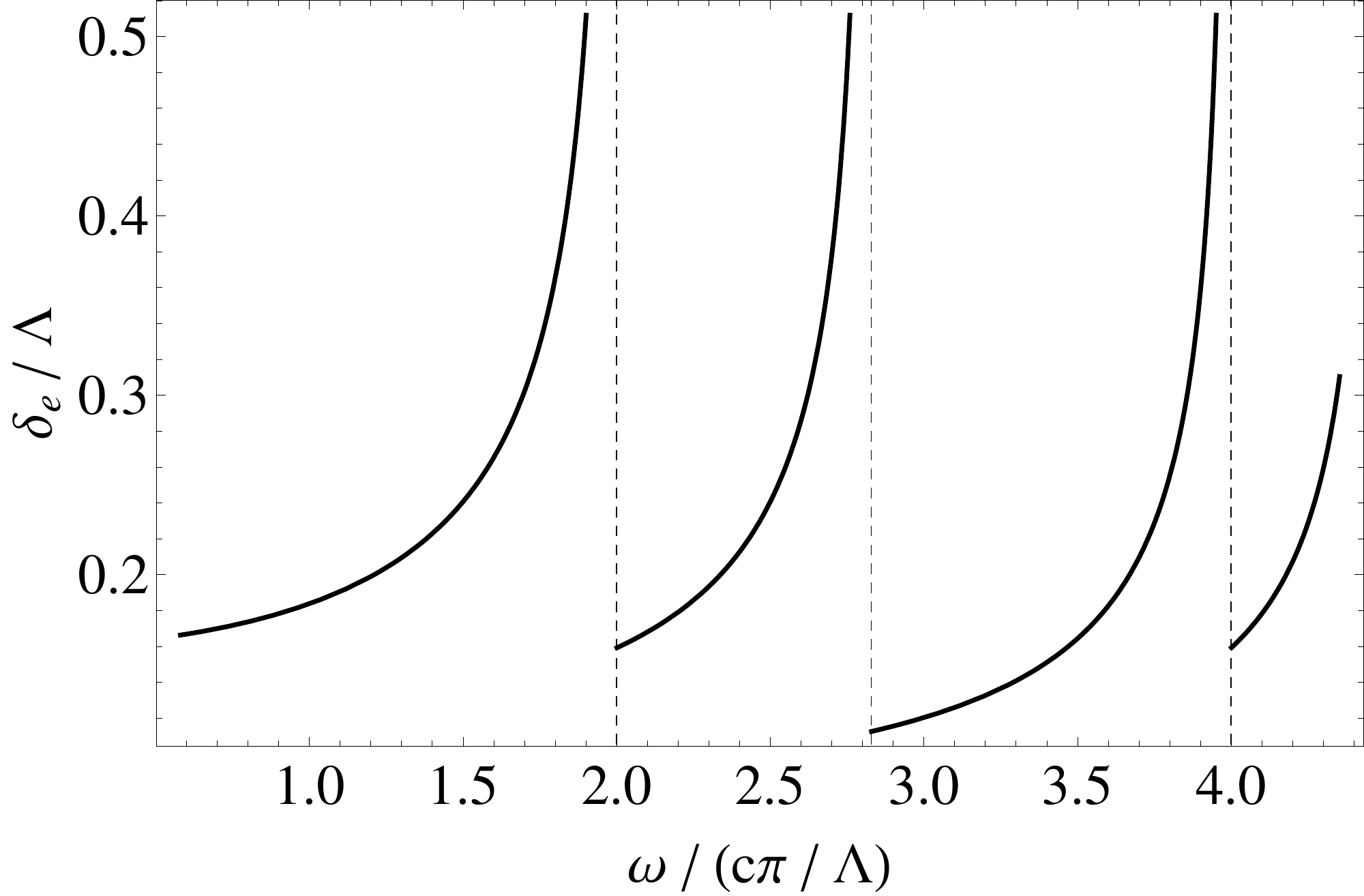}
    \caption{We plot the cavity field leakage length $\delta_e$ as a function of normalized frequency. Discontinuities at dashed vertical lines are the locations of the onset of propagating diffracted modes.}
    \label{fig:CMextend}
\end{figure}

There is one notable limitation of this approach, which we now discuss. At a frequency where the radicand in Eq.~\eqref{eq:kzFq} is zero, i.e.\ a propagating diffraction frequency,
\begin{equation}
\omega_{\textrm{diff}}=\frac{c \pi}{p \sqrt{\epsilon_s}} \sqrt{m^2+n^2},
\end{equation}
the field leakage depth asymptotically approaches infinity, with a discontinuity at the diffraction frequency. This can be seen in Fig.~\ref{fig:CMextend}, which shows the leakage depth as a function of frequency. The tendency towards infinite evanescent decay lengths captures the smooth transition from a localized mode to a propagating diffracted mode. However, approaching this transition makes the lowest order evanescent fields less local, while the second order (shorter decay length) evanescent fields are still localized. The actual effective cavity addition $\delta_e$ must then depend on the relative weights of the first- and second-order evanescent modes, which we do not analyze. Thus, we expect this model to break down at frequencies just below the diffraction frequencies. It is likely that this analysis can be improved by considering the scattering efficiencies into different diffracted modes, however even without that addition, this model is still highly accurate for nearly all frequencies.

Until this point we have been analyzing open apertures in metal films. We feel it is also worthwhile briefly noting that this approach can be generalized to closed cavities, with a different effective height,
\begin{equation}
\label{eq:effHhole}
h_{\textrm{eff}}=h+\delta_e+\delta_m,
\end{equation}
where $\delta_m$ is the field penetration depth into the metal, and where we follow the assumption of our previous work, neglecting fields beyond a skin depth into metal.\cite{LanseyJOSAB2011}

Using these results, we can rewrite Eq.~\eqref{eq:resonanceNoSD} using the effective height of Eq.~\eqref{eq:effHapt},
\begin{equation}
\label{eq:resonanceYesSD}
\omega_{mnp}=\frac{c}{\sqrt\epsilon_c}\left[\left(\frac{p\,\pi}{h+2\delta_e(\omega_{mnp})}\right)^2+\beta_{mn}^2\right]^{1/2}.
\end{equation}
Since the effective height is itself a function of frequency, it is difficult to find an exact solution for $\omega$ in all cases. Nevertheless, due to the analytic nature of this expression, we are still able to extract general trends in resonance changes due to cavity geometry. For ease of interpretation, we utilize the graphical approach discussed earlier.

\section{ECR dependence on structure geometry}
\label{sec:geom}
\begin{figure}[H]
    \centering
\subfigure[]{\includegraphics[width=.65\linewidth]{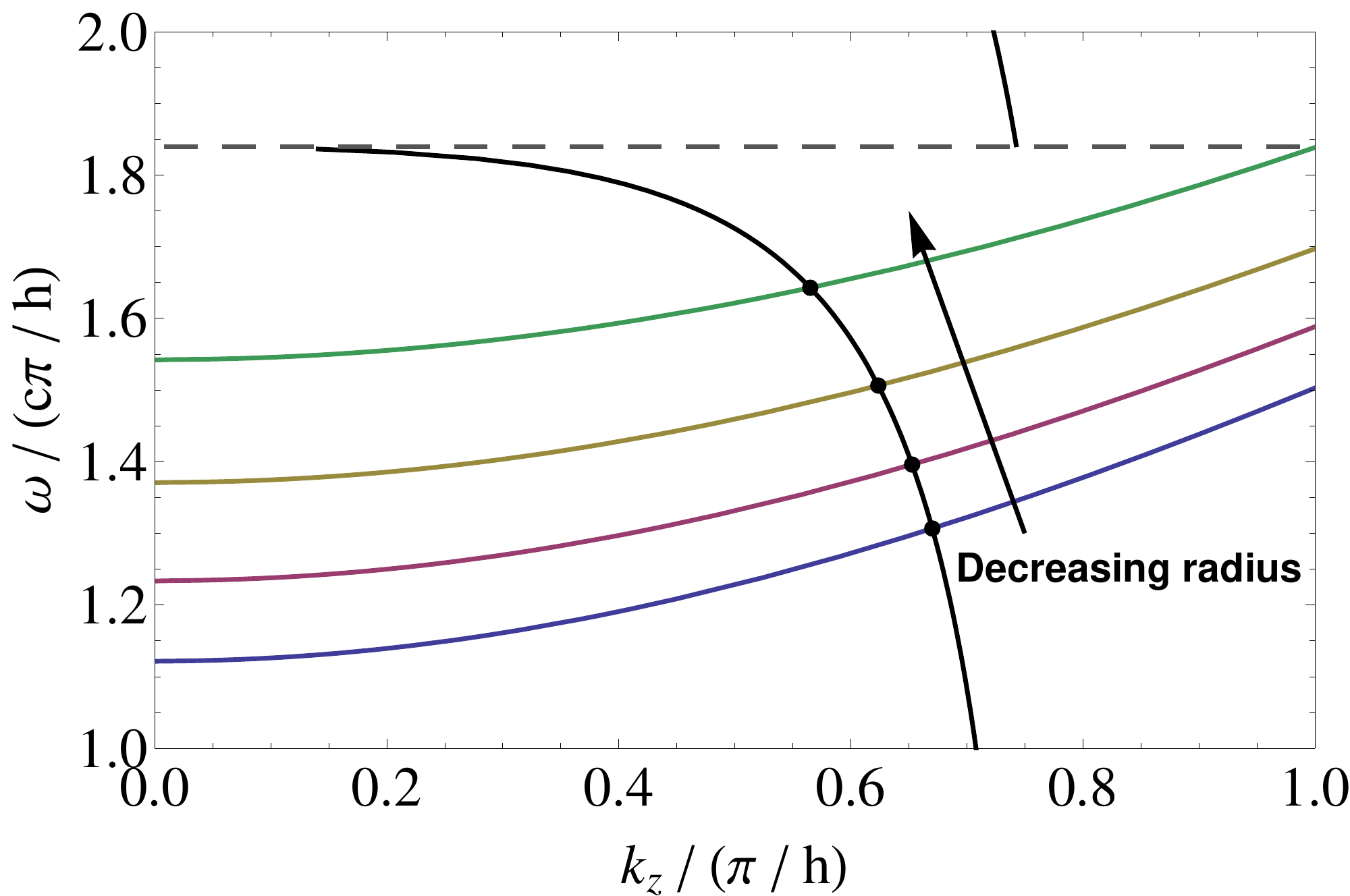}
    \label{fig:rvaryDisp}
    }
    \subfigure[]{\includegraphics[width=.65\linewidth]{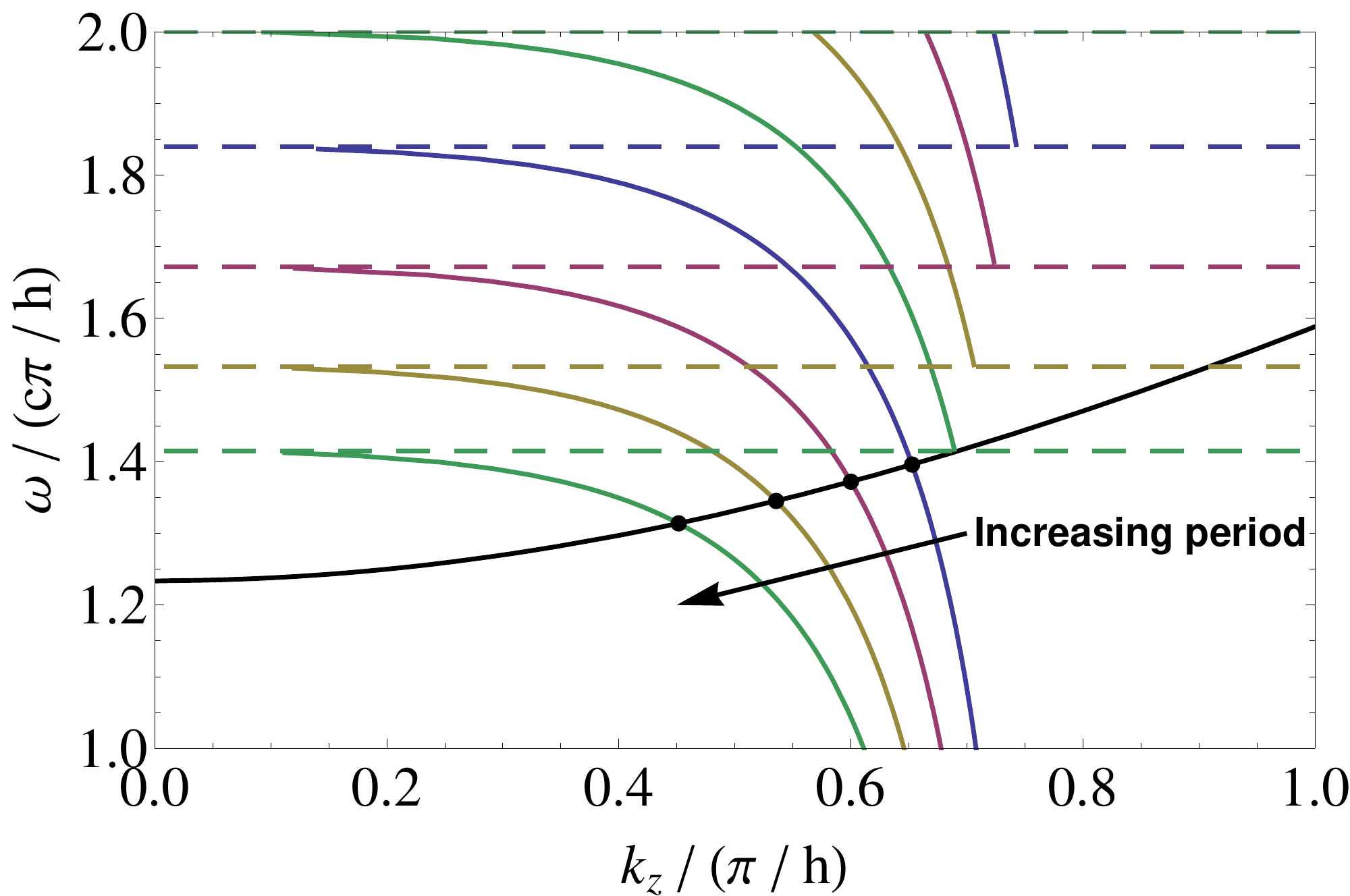}
    \label{fig:pvaryDisp}
    }
\\
\subfigure[]{\includegraphics[width=.65\linewidth]{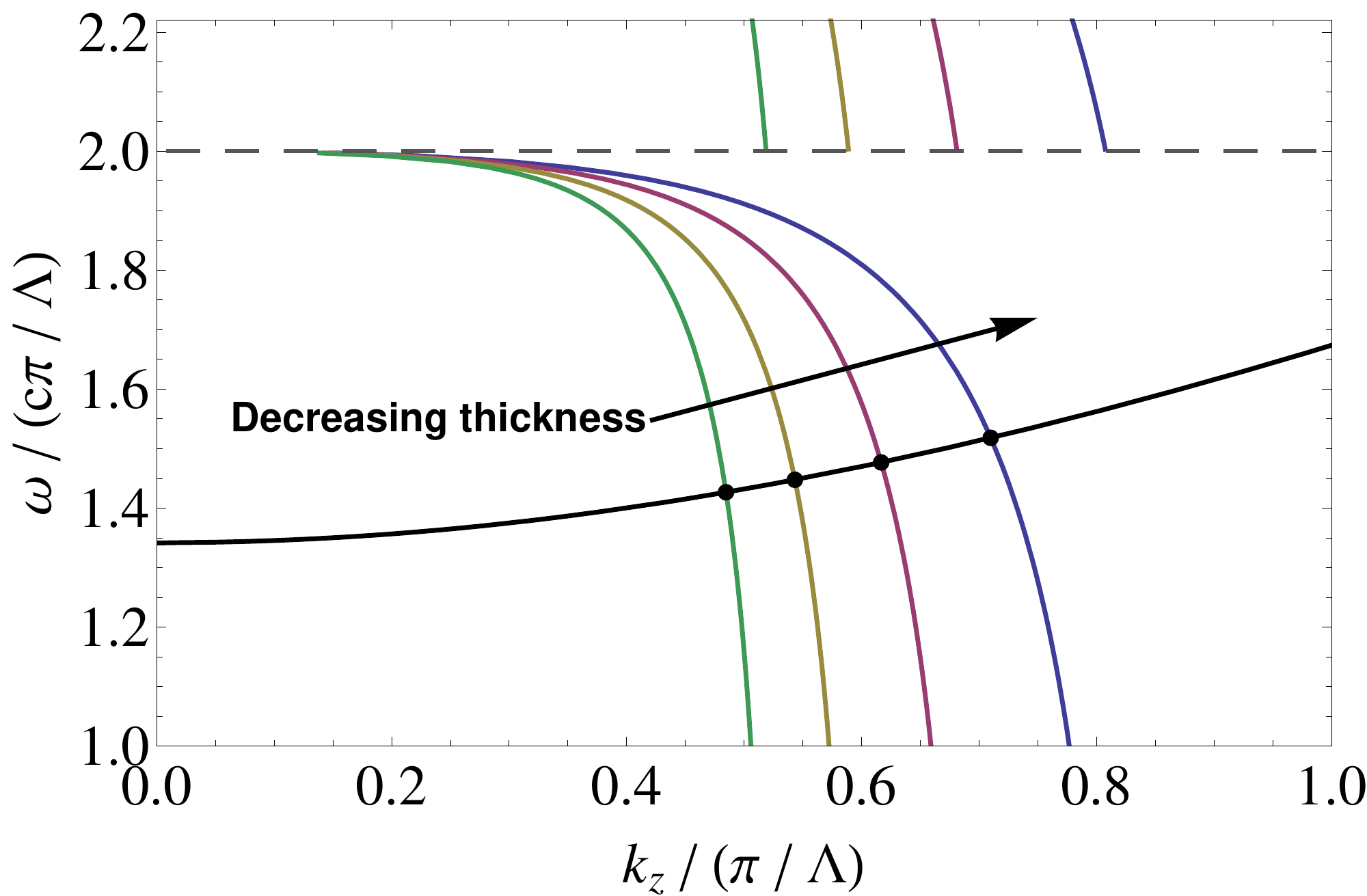}
    \label{fig:hvaryDisp}
    }
\caption{Graphical interpretation of the variation in the resonance condition due to changes in geometric parameters. \subref{fig:rvaryDisp} Effect of varying the radius. Arrow points along direction of decreasing radius with $2a/\Lambda$ of 0.7, 0.79, 0.87, and 0.96 plotted. Here $h/\Lambda=0.92$, the black line is the $k_z$ restriction from film thickness, and the colored lines are the different waveguide modal dispersion curves.\subref{fig:pvaryDisp}  Effect of varying the periodicity. Arrow points along direction of increasing periodicity with $p/h$ of 0.87, 0.96, 1.04, and 1.13 plotted. Horizontal dashed lines are the diffraction frequencies for each period. Here $h/a=2.63$, the black line is the waveguide modal dispersion curve, and the colored lines are the $k_z$ restrictions from film thickness. \subref{fig:hvaryDisp} Effect of varying the film thickness. Arrow points along direction of increasing thickness with $h/a$ of 2.11, 2.63, 3.16, and 6.32 plotted. Here $2a/\Lambda=0.87$, the black line is the waveguide modal dispersion curve, and the colored lines are the $k_z$ restrictions from film thickness.}
\label{fig:varyDisp}
\end{figure}

Figure~\ref{fig:varyDisp} shows the dispersion curves overlaid with the restricted $k_z$ condition (Eq.~\eqref{eq:kzYesSD}) for different geometries. Note the discontinuity in the $k_z$ restriction at diffraction frequencies due to the limitations of the theory discussed earlier. The dependence of EOT peaks on cavity radius is found in Eq.~\eqref{eq:resonanceYesSD}; the radius only affects the waveguide mode dispersion. Thus, keeping the film thickness and periodicity fixed, and changing the radius, shifts only the modal dispersion curve up or down, see Fig.~\ref{fig:rvaryDisp}. As the radius decreases, the allowed waveguide modes shift to higher frequencies, as expected. However, the rate of shifting is not uniform, in contrast to the FB model discussed in Section~\ref{sec:illus}, and smaller shifts are found for the same change in radius as the resonance approaches a diffraction frequency.

Changing the period or film thickness leaves the waveguide modal dispersion curve untouched, shifting only the restriction on $k_z$ due to the effective cavity height. The relative size of $h$ and $\delta_e$ determine the dominant contribution to the effective length. When $h$ is large relative to $\delta_e$, i.e.\ thick films, the resonance approaches the simpler model discussed earlier, where the dominant length is in the waveguide. When $\delta_e$ is large, i.e.\ extremely thin films or near diffraction, the effects of periodicity dominates the transmission spectrum.

Increasing the thickness of the film pushes the $k_z$ restriction curve closer to the straight vertical lines of Fig.~\ref{fig:kzSolveNoSD}. However, the lines never do reach $p\pi/h$ as there is always some field coupling depth, see Fig.~\ref{fig:hvaryDisp}. The smooth transition from the enhanced transmission of a periodic structure, to propagation along regular, independent waveguides can be seen.

Increasing the period shifts the first diffraction towards lower frequencies, drastically changing the curvature of this restriction curve, see Fig.~\ref{fig:pvaryDisp}. This, in turn, shifts all resonances towards lower frequencies, with the most drastic changes occurring close to diffraction frequencies. As the period increases further, it pushes down the diffraction frequencies, and thereby increases the density of restricted $k_z$ lines crossing the dispersion curves. It should also be noted that expected transmission through a film drops above a diffraction frequency. If the lowest order scattered diffracted mode is a propagating mode, it leaves proportionally less electromagnetic energy to be scattered into the lowest order evanescent mode, thus decreasing the available localized light which can be coupled to the ECR.

This result further explains the differences in EOT between periodic and single apertures. As the period approaches infinity (i.e.\ single apertures), the discrete transverse diffracted wavevectors $\sqrt{(m^2+n^2)K^2}$ can be made arbitrarily close together and can be considered a continuous variable which varies smoothly between 0 and $\epsilon_s\omega^2/c^2$. This leads to an infinite continuum of propagating reflected modes (i.e.\ a spherical scattered wave) and the number of evanescently decaying modes approaches zero. Thus, the number of modes which permit EOT approaches zero, leading to the expected weaker overall coupling and lower transmission.

\section{Comparison to simulation}
It is straightforward to numerically find the roots to Eq.~\eqref{eq:resonanceYesSD} and calculate the dependence of peaks in EOT on geometrical properties over a large range of values. To verify the predictions, we simulated structures using HFSS, which is a commercially available full-wave finite element simulation tool. We simulate periodic cylindrical apertures embedded in idealized PEC metal film as well as cavities in a realistic gold film.

\begin{figure}[H]
    \centering
\subfigure[Effect of varying the radius. Here $h/\Lambda=0.92$.]{\includegraphics[width=.65\linewidth]{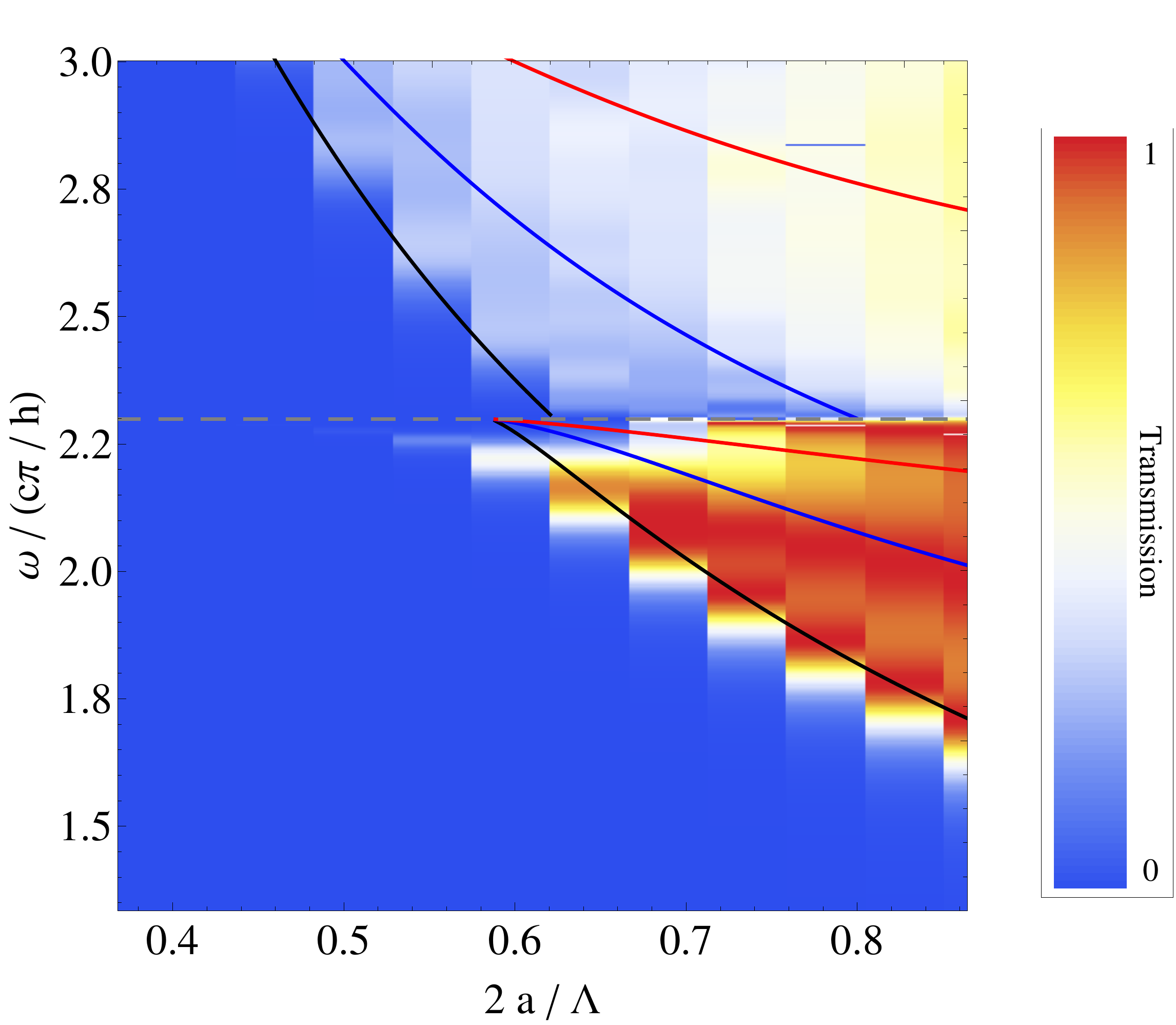}
    \label{fig:rvary}
    }
    \subfigure[Effect of varying the periodicity. Here $h/a=2.63$.]{\includegraphics[width=.65\linewidth]{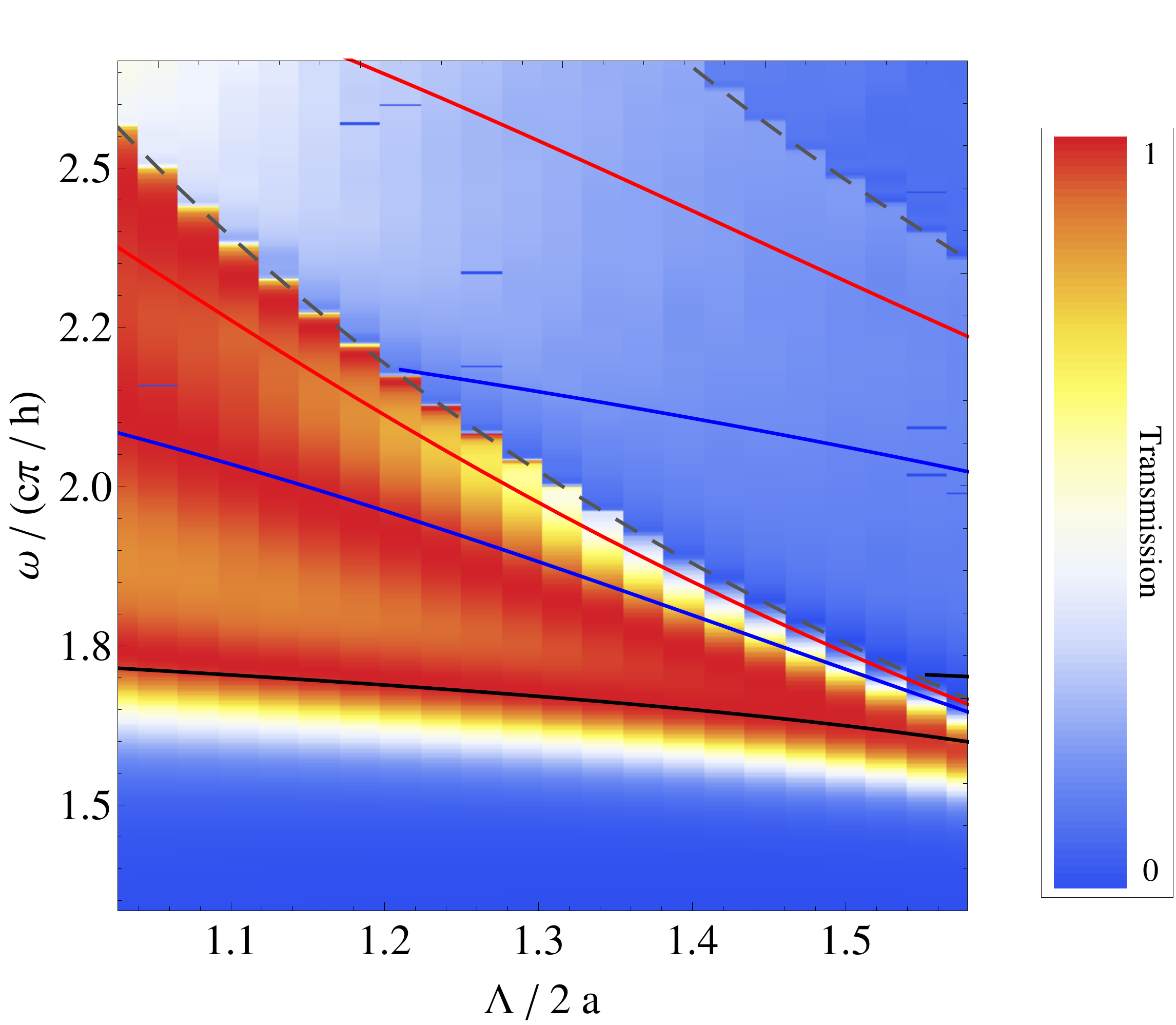}
    \label{fig:pvary}
    }
\\
\subfigure[Effect of varying the film thickness. Here $2a/\Lambda=0.87$.]{\includegraphics[width=.65\linewidth]{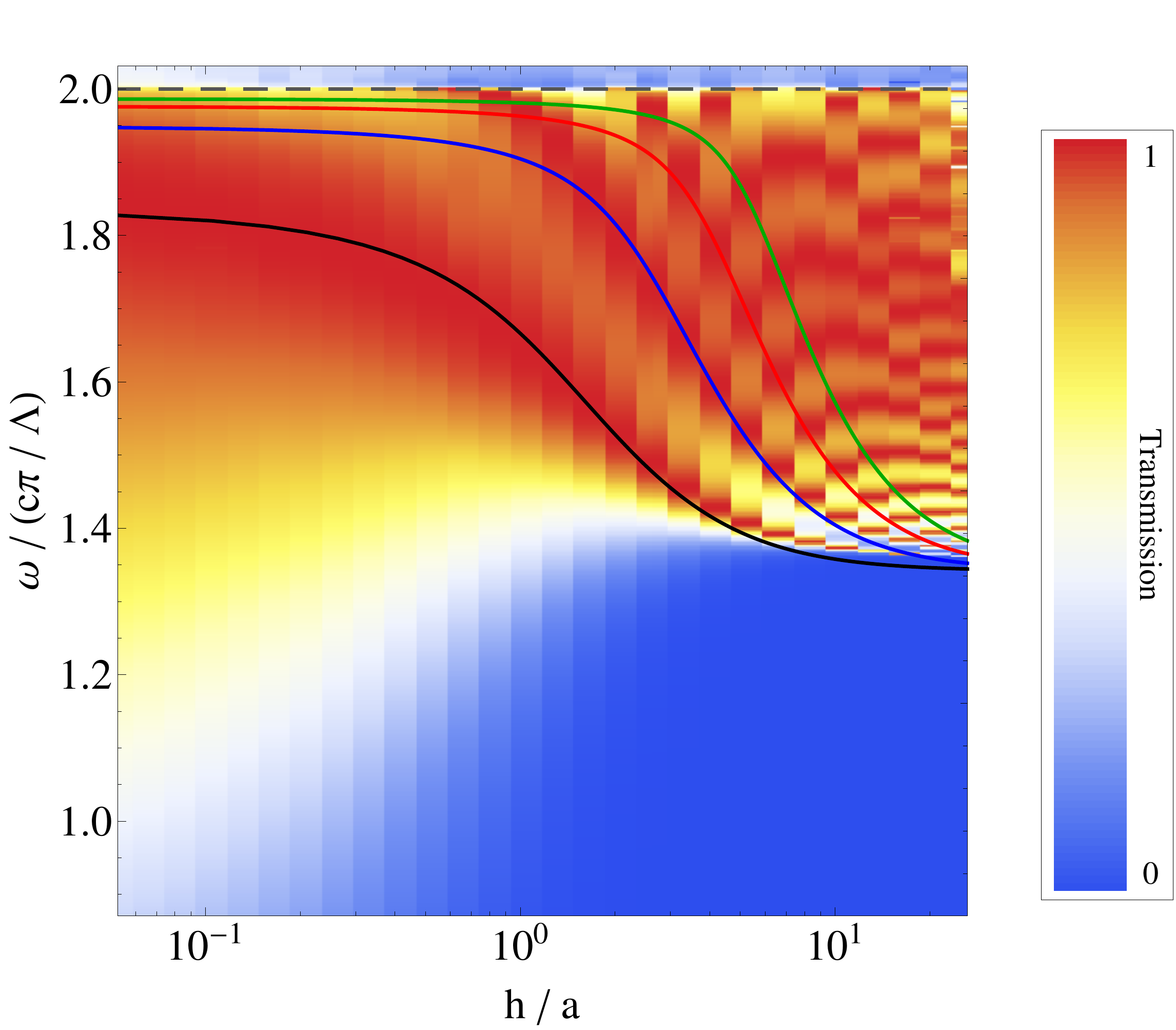}
    \label{fig:hvary}
    }
\caption{Simulated zero-order transmission through cylindrical apertures embedded in PEC, overlayed with predicted ECR peaks. The gray dashed lines are diffraction frequencies, and the black, blue and red (and green) colored lines are the lowest order $\mathrm{TE}_{11p}$ curves for p=1,2,3 (and 4) respectively as a function of radius \subref{fig:rvary}, periodicity \subref{fig:pvary} and film thickness \subref{fig:hvary}.}
\label{fig:vary}
\end{figure}

Figure~\ref{fig:vary} compares the simulated transmission through a PEC film with $\epsilon_c=\epsilon_s=1$, overlayed with predicted peaks of EOT. There is extremely strong agreement between the predicted and simulated results. Any major differences between predicted and simulated values occur at frequencies very close to diffraction frequencies, which is a manifestation of the limitation of this theory discussed earlier.

Note, that the subwavelength condition $2a/\lambda<1$ is given in the normalized coordinates of Fig.~\ref{fig:rvary} by
\begin{equation}
\left(\frac{2a}{\Lambda}\right) \left(\frac{\omega}{c\pi/h}\right)
<2\left(\frac{h}{\Lambda}\right).
\end{equation}
Similarly, in the normalized coordinates of Fig.~\ref{fig:pvary} by
\begin{equation}
\left(\frac{\omega}{c\pi/h}\right)
<\left(\frac{h}{a}\right),
\end{equation}
and in the normalized coordinates of Fig.~\ref{fig:hvary} by
\begin{equation}
\left(\frac{\omega}{c\pi/\Lambda}\right)
<2\left(\frac{2a}{\Lambda}\right)^{-1},
\end{equation}
Then, the plotted range of values in Fig.~\ref{fig:vary} are entirely within the subwavelength regime. The existence of EOT through subwavelength PEC structures, where SP resonance is not a contributing factor, further highlights the minimal role of SPs in EOT. The ECR can be viewed as the dominant mechanism which enhances transmission for this class of structures.

\begin{figure}[h]
    \centering
\subfigure[Effect of varying the radius. Here $h=250nm$ and $\Lambda=800nm$.]{\includegraphics[width=.65\linewidth]{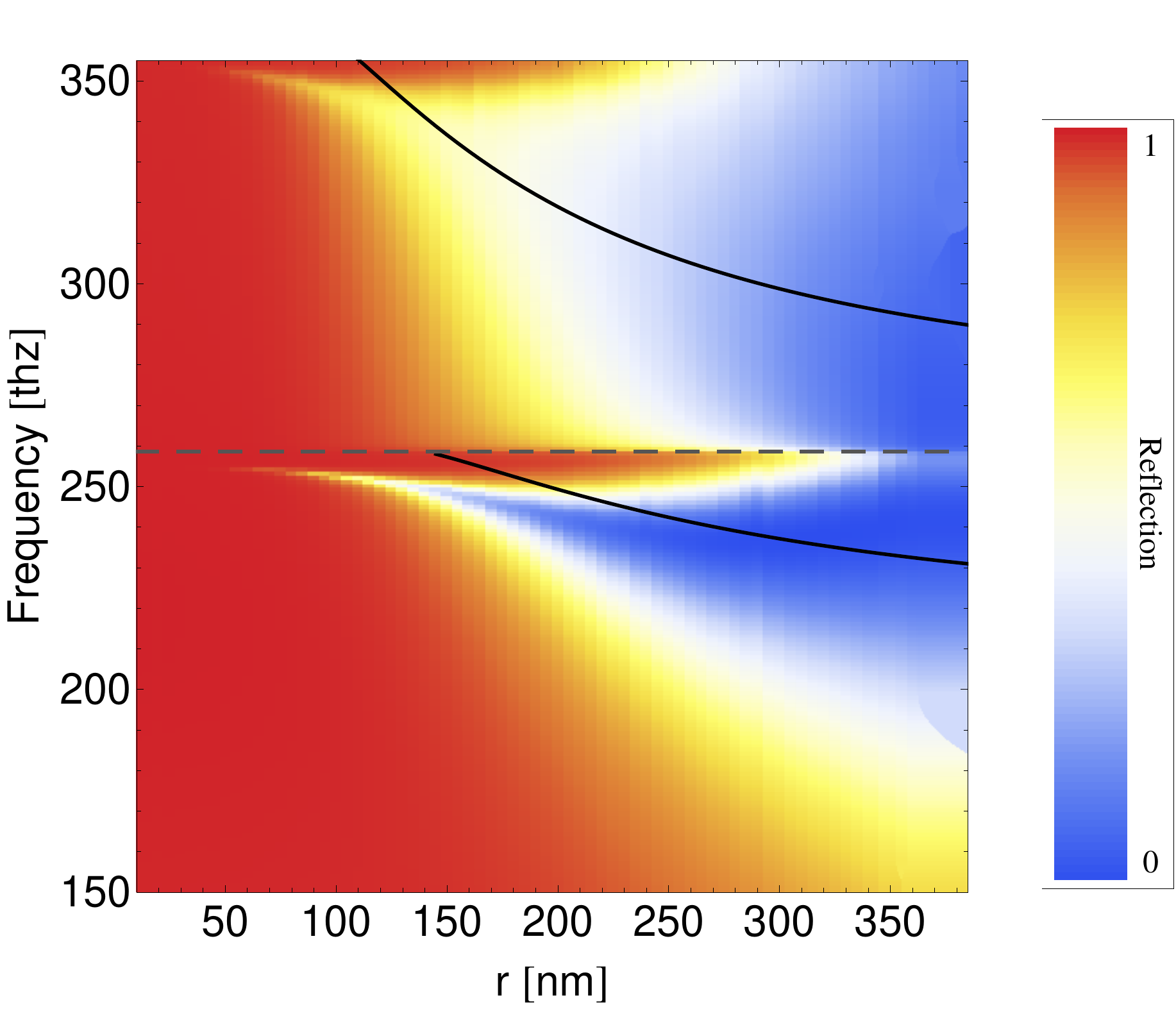}
    \label{fig:rvaryAu}
    }
    \subfigure[Effect of varying the periodicity. Here $h=250nm$ and $a=190nm$.]{\includegraphics[width=.65\linewidth]{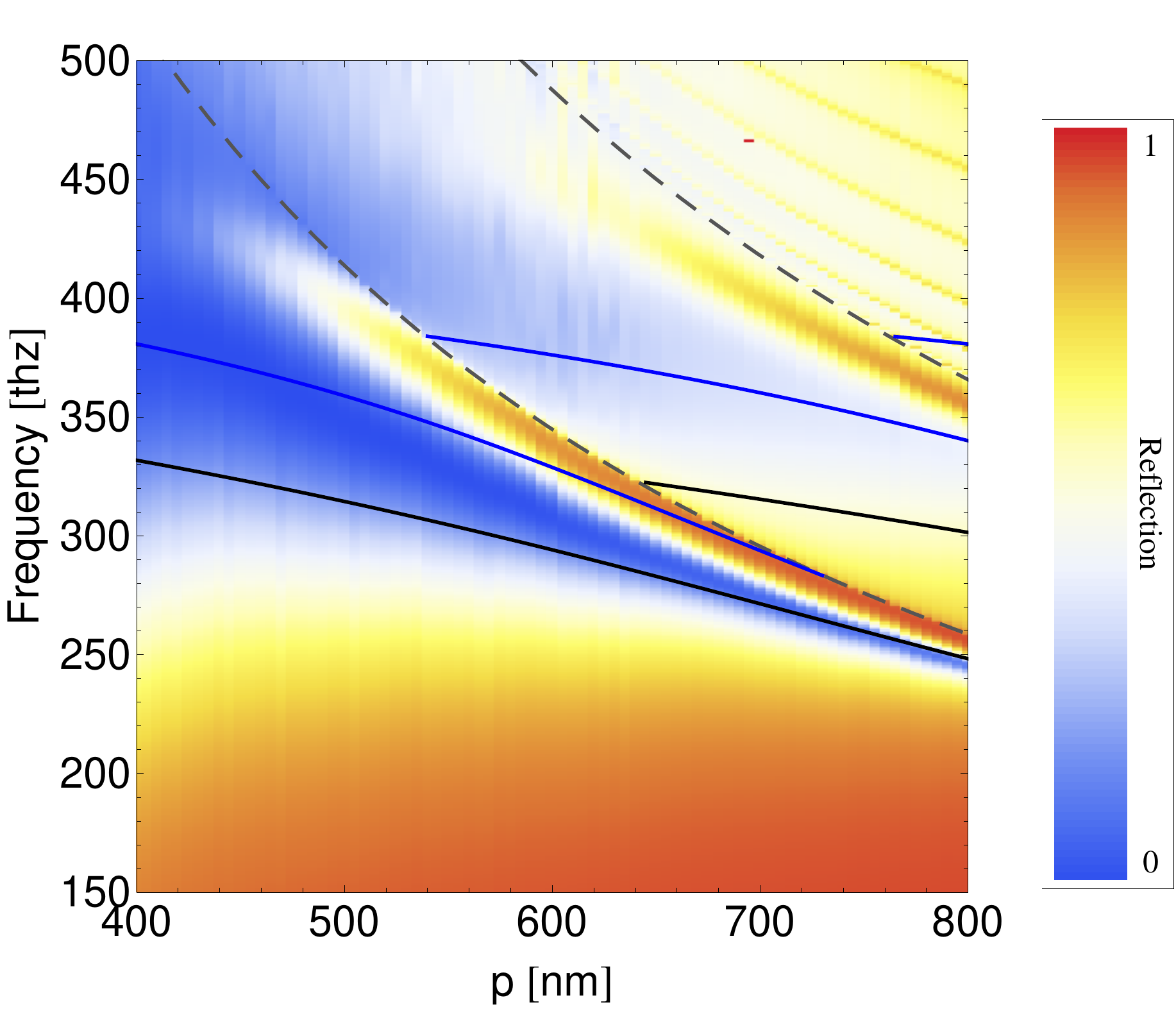}
    \label{fig:pvaryAu}
    }
\\
\subfigure[Effect of varying the film thickness. Here $a=190nm$ and $\Lambda=800nm$.]{\includegraphics[width=.65\linewidth]{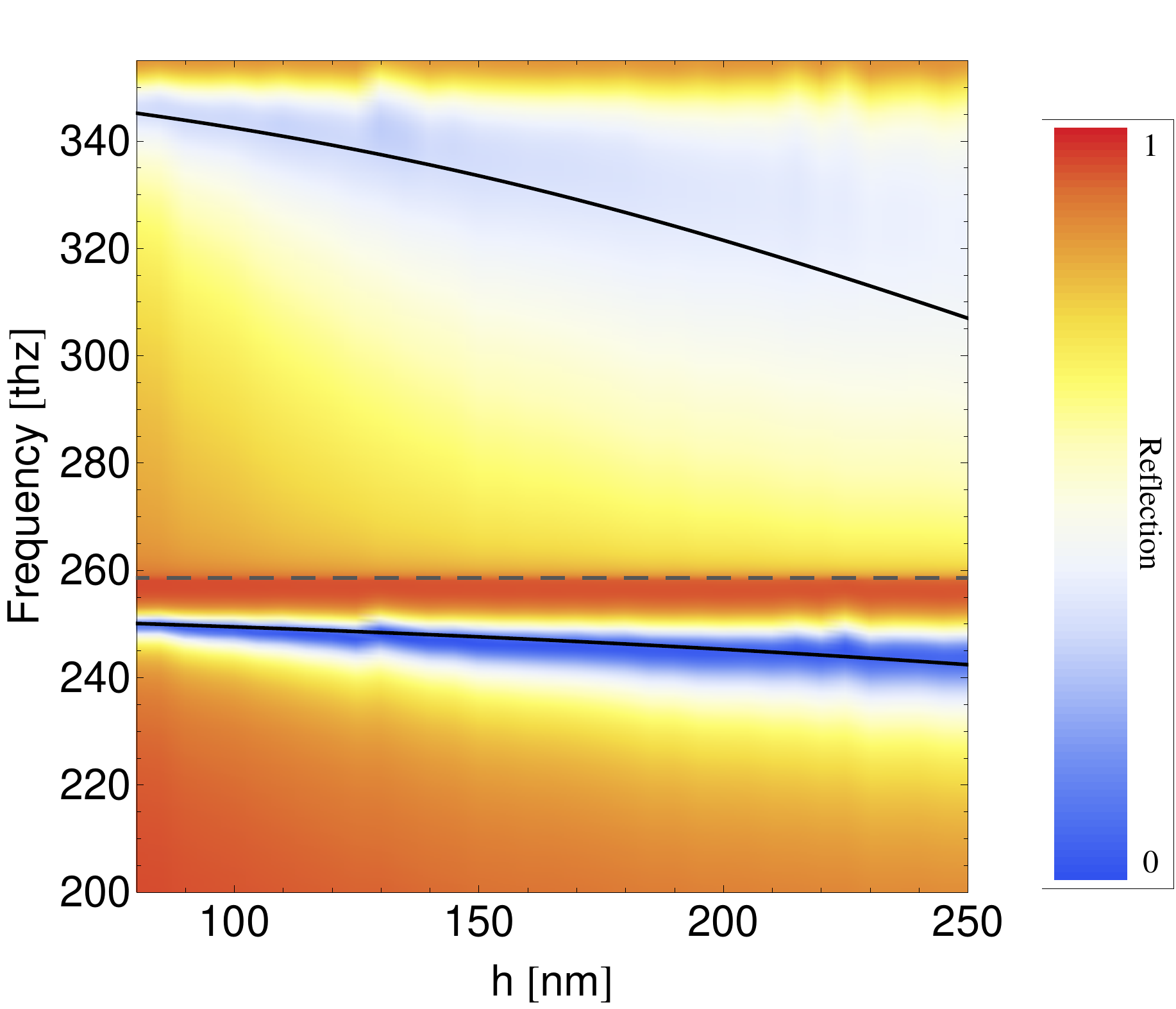}
    \label{fig:hvaryAu}
    }
\caption{Simulated specular reflection from an array of cylindrical holes embedded in gold, overlayed with predicted ECR peaks. The gray dashed lines are diffraction frequencies, the black and blue lines are the lowest order $\mathrm{TE}_{111}$ and $\mathrm{TM}_{011}$ curves, respectively as a function of radius \subref{fig:rvaryAu}, periodicity \subref{fig:pvaryAu} and film thickness \subref{fig:hvaryAu}.}
\label{fig:varyAu}
\end{figure}

As further verification of the validity of this approach, we compare the predictions of this model as it applies to cavities in metal, as per Eq.~\eqref{eq:effHhole}. Figure~\ref{fig:varyAu} compares the simulated specular reflection from a gold film with $\epsilon_s=2.1$ and $\epsilon_c=2.1 + i\,0.9$, overlayed with predicted peaks of EOT. We introduced loss in the cavity dielectric to better identify the resonances; when there are strong fields built up in the cavity there is increased loss in the dielectric. There is reasonably close agreement between the predicted and simulated results. The weakest agreement is found upon varying the radius. Due to the fairly large skin depths of gold at optical frequencies, there is a small range of values where $a/\delta_m$ is large, while $2a/\Lambda$ remains small. When these conditions are not satisfied, a large fraction of the fields are contained within the metal, and the ECR model and SDBC would not apply. Any other differences between predicted and simulated values occur at frequencies very close to diffraction frequencies, which is a manifestation of the limitations of this theory discussed earlier in Section~\ref{sec:ecr}.

\section{Calculating the quality factor of the ECR}
Now that we are able to calculate the resonance frequencies of the effective cavities, it is possible to calculate the quality factor, $Q$, of the effective resonance. Here we use the expression,
\begin{equation}
\label{eq:qthy}
Q=\omega_{mnp}\frac{U}{P_{\textrm{loss}}},
\end{equation}
where $\omega_{mnp}$ is a particular resonance frequency found using Eq.~\eqref{eq:resonanceYesSD}, $U$ is the time-averaged energy stored within the aperture and $P_{\textrm{loss}}$ is the time-average power lost.\cite{Jackson2} Using the known field expressions,\cite{LanseyJOSAB2011}
\begin{equation}
\psi_{mnp}=e^{i(p\pi/h_{\textrm{eff}})z}e^{im\varphi}J_m\left(\beta_{mn}r\right),
\end{equation}
we can calculate both the energy stored and power lost.

The energy stored in the aperture is

\begin{equation}
\label{eq:Ustored}
U_{mnp}=\frac{C}{4\pi}\left[\epsilon \frac{\omega_{mnp}^2}{c^2}\frac{1}{ \beta_{mn}^2}\right]\frac{h}{4}\iint\left|\psi_{mnp}\right|^2rdr\,d\varphi
\end{equation}
where the integral is taken over the aperture face, and where $C=1$ for TE modes and $C=\epsilon$ for TM modes.\cite{Jackson2}

The power lost from the aperture is due to two sources: transmission out of the bottom face of the aperture and losses at the metal walls, the latter of which only exists with non-PEC metals. We do not consider the loss from the upper face of the aperture. The reason for this is apparent from the PEC case where we find 100\% transmission through the film at resonance. This means that an equal amount of power flows through the top face of the apertures as through the bottom, with the same directionality. Since the unit normal vectors pointing out of the cavity are equal in magnitude but opposite in direction at the top an bottom faces, the total power flow \emph{out} of the cavity is zero. This would correspond to an infinite $Q$, which must be rejected outright. Thus, we assume that, at resonance, whatever power enters the aperture at the top face must exit from the bottom or be absorbed in the walls.

This result allows us to greatly simplify the calculation for $P_{\textrm{loss}}$, as we need not directly solve for the flow of power into the metal walls. Given that all energy which enters the apertures from the top surface must either be absorbed in the metal or transmitted out of the aperture, the power lost must be equal to the power incident, $P$, and it is sufficient to set $P_{\textrm{loss}}=P$. The incident power is,\cite{Jackson2}
\begin{multline}
\label{eq:Pin}
P_{mnp}=\frac{C}{4\pi}\frac{c}{2\sqrt{\epsilon }}\left[\epsilon \frac{\omega_{mnp}^2}{c^2}\frac{1}{ \beta_{mn}^2}\right]\\
\times\sqrt{1-\left[\epsilon \frac{\omega_{mnp}^2}{c^2}\frac{1}{ \beta_{mn}^2}\right]^{-1}}
\iint\left|\psi_{mnp}\right|^2rdr\,d\varphi
\end{multline}
which is also the power lost.

Substituting Eqs.~\eqref{eq:Ustored} and \eqref{eq:Pin} into Eq.~\eqref{eq:qthy} gives an expression for the $Q$ of a particular resonance,

\begin{equation}
\label{eq:Q}
Q_{mnp}=\frac{h h_{\text{eff}}}{2}\frac{\epsilon   }{p\pi}\frac{\omega_{mnp}^2}{c^2}.
\end{equation}
Note that $Q$ is proportional to the resonance frequency squared. Thus, for an otherwise identical structure, transitioning from a PEC to a real metal decreases the resonance frequency, and thus decreases the Q which implicitly captures the metallic losses. This, furthermore, captures the dependence of the quality factor on the various geometrical parameters.

\section{Summary and conclusion}
We have presented a new analytical theory for the mechanism of EOT through arrays of subwavelength apertures. This theory demonstrates that an effective cavity is described by the cavity dimensions in conjunction with the decay length of strong, localized evanescent diffracted modes in the regions above and below the metal film. These localized fields are the primary cause for coupling light between the apertures and the other regions. Thus, this model is a fundamental theory for the mechanism of EOT.

Furthermore, we have shown how to predict the frequencies where peaks in enhanced transmission occur and how these frequencies depend on the cavity dimensions, metal choice, as well as periodicity of the structure. This model is valid over an extremely broad range of geometries, limited only at frequencies close to diffraction. We have shown strong agreement between our theory and simulations for both apertures through PEC and cavities embedded in real gold.

Although the model was applied to cylindrical apertures, this approach can be generalized to apertures of arbitrary shape. Other apertures change the value of the transverse wavevector $\beta$ in dispersion relation of Eq.~\eqref{eq:disp}, leaving everything else unchanged. Similarly, this can be generalized to rectangular, and perhaps arbitrary, periods, changing only the reciprocal lattice vectors in Eq.~\eqref{eq:ksup}. Likewise, the dependence of EOT on incident angle can likely be deduced from the changes this makes to the propagation vectors in the superstrate.

\begin{acknowledgments}
This material is based upon work supported by the DOD/DARPA SBIR PROGRAM under Contract No. W31P4Q-11-C-0238.  The views, opinions, and/or findings contained in this article/presentation are those of the author/presenter and should not be interpreted as representing the official views or policies, either expressed or implied, of the Defense Advanced Research Projects Agency or the Department of Defense. 

Computational support for this work was provided by the CUNY Center for Advanced Technology's high performance computing cluster.
\end{acknowledgments}


%

\end{document}